\magnification \magstep1
\raggedbottom
\openup 2\jot
\voffset6truemm
\def\cstok#1{\leavevmode\thinspace\hbox{\vrule\vtop{\vbox{\hrule\kern1pt
\hbox{\vphantom{\tt/}\thinspace{\tt#1}\thinspace}}
\kern1pt\hrule}\vrule}\thinspace}
\centerline {\bf SCATTERING FROM SINGULAR POTENTIALS}
\centerline {\bf IN QUANTUM MECHANICS}
\vskip 1cm
\leftline {Giampiero Esposito$^{1,2}$}
\vskip 1cm
\noindent
${ }^{1}${\it Istituto Nazionale di Fisica Nucleare, Sezione
di Napoli, Complesso Universitario di Monte S. Angelo, Via
Cintia, Edificio G, 80126 Napoli, Italy}
\vskip 0.3cm
\noindent
${ }^{2}${\it Universit\`a di Napoli Federico II, Dipartimento
di Scienze Fisiche, Mostra d'Oltremare Padiglione 19,
80125 Napoli, Italy}
\vskip 1cm
\noindent
{\bf Abstract}.
In non-relativistic quantum mechanics, singular potentials in
problems with spherical symmetry lead to a Schr\"{o}dinger 
equation for stationary states with non-Fuchsian singularities
both as $r \rightarrow 0$ and as $r \rightarrow \infty$. In the
sixties, an analytic approach was developed for
the investigation of scattering from such potentials,
with emphasis on the polydromy of the wave function in the
$r$-variable. The present paper extends those early results to
an arbitrary number of spatial dimensions. The 
Hill-type equation which leads, in principle, to the evaluation
of the polydromy parameter, is obtained from the Hill equation
for a two-dimensional problem
by means of a simple change of variables.
The asymptotic forms of the wave function as $r \rightarrow 0$
and as $r \rightarrow \infty$ are also derived. The Darboux
technique of intertwining operators is then applied to obtain
an algorithm that makes it possible to solve the Schr\"{o}dinger
equation with a singular potential containing many negative 
powers of $r$, if the exact solution with even just one term
is already known.
\vskip 100cm
\leftline {\bf 1. Introduction}
\vskip 0.3cm
\noindent
One of the longstanding problems of non-relativistic quantum
mechanics is the investigation of scattering from singular
potentials, with efforts by many authors over several decades
(see [1--15] and references therein). The main motivations 
can be described as follows [13].
\vskip 0.3cm
\noindent
(i) Repulsive singular potentials make it possible to obtain a
fairly accurate description of the short-range part of the
nucleon-nucleon interaction.
\vskip 0.3cm
\noindent
(ii) The $(p,p)$ and $(p,\pi)$ processes can be interpreted in
terms of complex potentials $r^{-n}$, with $n>2$.
\vskip 0.3cm
\noindent
(iii) Repulsive singular potentials reproduce also the interactions
of nucleons with $K$-mesons, and $\alpha-\alpha$ scattering
processes.
\vskip 0.3cm
\noindent
(iv) The Lennard--Jones potential, proportional to $r^{-12}$, can
be used to study interactions among the overlapping electron
clouds of non-polar molecules.
\vskip 0.3cm
\noindent
(v) At a field-theoretical level, it appears quite remarkable that
non-renormalizable field theories give rise to effective potentials
in the Bethe--Salpeter equation which are singular [1, 2],
whereas super-renormalizable and renormalizable field theories 
give rise to regular or transition-type effective potentials,
respectively. There was therefore the hope that any new insight
gained into the analysis of non-relativistic potential scattering
in the singular case, could be eventually used to obtain a better
understanding of quantum field theories for which perturbative
renormalization fails (cf section 6).
\vskip 0.3cm
\noindent
(vi) In particular, one might then hope to be able to ``map"
the analysis of quantum gravity based on the Einstein--Hilbert
action (plus boundary terms), which is well known to be
incompatible with the requirement of perturbative
renormalizability [16], into a scattering problem in the singular
case, for which the Schr\"{o}dinger equation for stationary
states:
$$
\left[{d^{2}\over dr^{2}}+{(q-1)\over r}{d\over dr}
-{l(l+q-2)\over r^{2}}+k^{2}\right]\psi(r)
=V(r)\psi(r)
\eqno (1.1)
$$
has non-Fuchsian singularities (see appendix) 
both as $r \rightarrow 0$ and 
as $r \rightarrow \infty$. With our notation, $q$ is the number
of spatial dimensions, and $l(l+q-2)$ is obtained by studying
the action of the Laplace--Beltrami operator on wave
functions belonging to the tensor product [17] 
$$
L^{2}({\cal R}_{+},r^{q-1}dr) \otimes 
L^{2}(S^{q-1},d\Omega).
$$
Moreover, with a standard notation, 
one has (of course, the energy $E$ is positive in a
scattering problem)
$$
k^{2} \equiv {2mE \over {\hbar}^{2}}
\eqno (1.2)
$$
$$
V(r) \equiv {2m \over {\hbar}^{2}}U(r)
\eqno (1.3)
$$
with $U(r)$ the potential term in the original form of the
Schr\"{o}dinger equation. From now on, it is $V(r)$ which will
be referred to as {\it the potential}, following the 
convention in the literature.

Among the analytic results obtained so far in the investigation 
of potential scattering in the singular case, we find it 
appropriate to mention what follows.
\vskip 0.3cm
\noindent
(i) A constructive determination of the $S$-matrix, based on the
polydromy properties of the wave function 
(see appendix) and on the Hill
equation for the polydromy parameter [6, 8].
\vskip 0.3cm
\noindent
(ii) Perturbative technique for the potential 
$V(r) \equiv g^{2}r^{-4}$ in three dimensions, by re-expressing the
radial Schr\"{o}dinger equation as a modified Mathieu equation [10],
with evaluation of $S$-matrix and Regge poles.
\vskip 0.3cm
\noindent
(iii) Generalized variable-phase approach, leading to a JWKB 
phase-shift formula [13]. 
\vskip 0.3cm
\noindent
(iv) Generalization of the JWKB method to arbitrary order, with
rigorous error bounds [14].

In our paper, sections 2 and 3 apply the method of [6, 8] to
the Schr\"{o}dinger equation for stationary states in three or
more spatial dimensions, proving that a simple but deep relation
exists between the corresponding Hill equations 
in two and three or more
spatial dimensions. Section 4 presents, for completeness, the
JWKB analysis of the wave function, jointly with its limiting
behaviour as $r \rightarrow 0$ and as $r \rightarrow \infty$.
Section 5 studies the application of the intertwining operator
technique to singular potential scattering. Results and open
problems are described in section 6.
\vskip 0.3cm
\leftline {\bf 2. Schr\"{o}dinger equation for stationary states}
\vskip 0.3cm
\noindent
Following the remarks in the introduction, we first study the
Schr\"{o}dinger equation for stationary states in three spatial
dimensions in a central potential:
$$
\left[{d^{2}\over dr^{2}}+{2\over r}{d\over dr}
-{l(l+1)\over r^{2}}+k^{2}\right]\psi(r)=V(r)\psi(r).
\eqno (2.1)
$$
What is crucial is the polydromy of the wave function in the
$r$ variable. Indeed, if the potential $V(r)$ is a single-valued
function of $r$, one can find two independent solutions
$$
\psi_{1}(r)=r^{\gamma} \; \chi_{1}(r)
\eqno (2.2)
$$
$$
\psi_{2}(r)=r^{-\gamma} \; \chi_{2}(r)
\eqno (2.3)
$$
where $\chi_{1}$ and $\chi_{2}$ are single-valued functions
of $r$, and $\gamma$ is a parameter which can be determined from
a transcendental equation (see below). The general solution of
Eq. (2.1) is therefore of the form
$$
\psi(r)=\alpha_{1} \; \psi_{1}(r)+\alpha_{2} \; \psi_{2}(r).
\eqno (2.4)
$$
Remarkably, one can compute directly $\chi_{1}(r)$ and $\chi_{2}(r)$
and study their behaviour as $r \rightarrow 0$ and as 
$r \rightarrow \infty$ [6, 8]. For this purpose, the following
Laurent expansions are used (the subscript for $\chi$ is omitted
for simplicity):
$$
W(r) \equiv r^{2}[V(r)-k^{2}]=\sum_{n=-\infty}^{\infty}
w_{n}r^{n} \; \; r \in ]0,\infty[
\eqno (2.5)
$$
$$
\chi(r)=\sum_{n=-\infty}^{\infty}c_{n}r^{n} \; \; 
r \in ]0,\infty[.
\eqno (2.6)
$$
These expansions hold because $V(r)$ is assumed to be an
analytic function in the complex-$r$ plane, with singularities
only at infinity and at the origin [6, 8]. The Laurent series
(2.5) and (2.6) are now inserted into Eq. (2.1), which is 
equivalent to the differential equation (cf [6, 8])
$$
r^{2}{d^{2}\chi \over dr^{2}}
+2(\gamma +1)r{d\chi \over dr}
+\Bigr(\gamma(\gamma+1)-l(l+1) \Bigr)\chi
=r^{2}\Bigr[V(r)-k^{2}\Bigr]\chi.
\eqno (2.7)
$$
One thus finds the following infinite system of equations 
for the coefficients (cf [8]):
$$
\Bigr[(n+\gamma)(n+\gamma+1)-{\overline \lambda}^{2}
\Bigr]c_{n}=\sum_{m=-\infty}^{\infty}
{\overline u}_{n-m}c_{m}
\eqno (2.8)
$$
where
$$
{\overline \lambda}^{2}=l(l+1)+w_{0}
\eqno (2.9)
$$
$$
{\overline u}_{n}=w_{n}-w_{0}\delta_{n,0}.
\eqno (2.10)
$$
To solve the system (2.8) one first writes an equivalent system
for which the determinant of the matrix of coefficients is
well defined. Such a new system is obtained from (2.8) by
dividing the $n$-th equation by 
$(n+\gamma)(n+\gamma+1)-{\overline \lambda}^{2}$. The resulting 
matrix of coefficients has elements
$$
H_{n,m}=\delta_{n,m}-{{\overline u}_{n-m}\over
[(n+\gamma)(n+\gamma+1)-{\overline \lambda}^{2}]}
\eqno (2.11)
$$
where ${\rm det}(H)$ exists since the double series
$$
\sum_{n,m}{{\overline u}_{n-m}\over 
[(n+\gamma)(n+\gamma+1)-{\overline \lambda}^{2}]}
$$
converges for all values of $\gamma$ which do not correspond to
zeros of the denominator. At this stage one can appreciate the
substantial difference between regular and singular potentials.
In the former case, $u_{n}$ is non-vanishing only for positive $n$.
In the singular case, however, the presence of negative powers
in the Laurent series (2.5) gives $\gamma$ as the solution of a
transcendental equation, i.e. (the vanishing of ${\rm det}(H)$
being necessary and sufficient 
to find non-trivial solutions of the system (2.8))
$$
F(\gamma) \equiv {\rm det}(H)=0.
\eqno (2.12)
$$
\vskip 0.3cm
\leftline {\bf 3. Equation for the $\gamma$ parameter}
\vskip 0.3cm
\noindent
To evaluate $F(\gamma)$, we point out that, on defining
$$
{\widetilde \gamma} \equiv \gamma +{1\over 2}
\eqno (3.1)
$$
$$
{\widetilde \lambda}^{2} \equiv {\overline \lambda}^{2}
+{1\over 4}=\left(l+{1\over 2}\right)^{2}+w_{0}
\eqno (3.2)
$$
one finds
$$
(n+\gamma)(n+\gamma+1)-{\overline \lambda}^{2}
=(n+{\widetilde \gamma})^{2}-{\widetilde \lambda}^{2}.
\eqno (3.3)
$$
This simple but fundamental property makes it possible to
perform the three-dimensional analysis by relying entirely 
on the investigation in two spatial dimensions, because
$H_{n,m}$ now reads
$$
H_{n,m}=\delta_{n,m}-{{\overline u}_{n-m}\over
[(n+{\widetilde \gamma})^{2}-{\widetilde \lambda}^{2}]}
\eqno (3.4)
$$
and hence, from the work in [6, 8], one knows that
$$
F({\widetilde \gamma})=1+\Bigr[F(0)
-1 \Bigr]{\Bigr[\cot \pi(
{\widetilde \gamma}+{\widetilde \lambda})
-\cot \pi ({\widetilde \gamma}-
{\widetilde \lambda})\Bigr] \over
2 \cot \pi {\widetilde \lambda}}
\eqno (3.5)
$$
where $F$ is an even periodic function of $\widetilde \gamma$,
with unit period [8]. The equation
$$
F({\widetilde \gamma})=0
\eqno (3.6)
$$
is, as we said in section 2, a transcendental equation. If a root, 
say $x$, is known, at least approximately, 
one can then evaluate the desired $\gamma$ parameter from the
definition (3.1) as
$$
\gamma=x -{1\over 2}.
\eqno (3.7)
$$

The ground is now ready for understanding the key features of
singular potential scattering in an arbitrary number of spatial
dimensions. For this purpose, we remark that, upon setting
$\psi(r)=r^{\gamma}\chi(r)$, Eq. (1.1) leads to the following
second-order equation for $\chi$ (cf (2.7)):
$$ \eqalignno{
\; & \left[r^{2}{d^{2}\over dr^{2}}+(2\gamma+q-1)r{d\over dr}
\right . \cr
& \left . +\Bigr(\gamma^{2}+(q-2)\gamma-l(l+q-2)\Bigr)
\right]\chi(r)=W(r)\chi(r).
&(3.8)\cr}
$$
Thus, on defining (cf (3.1))
$$
{\widetilde \gamma} \equiv \gamma +{1\over 2}(q-2)
\eqno (3.9)
$$
one can re-express Eq. (3.8) in the form
$$ \eqalignno{
\; & \left[r^{2}{d^{2}\over dr^{2}}
+\Bigr(2{\widetilde \gamma}+1 \Bigr)r{d\over dr} \right . \cr
& \left . +\Bigr({\widetilde \gamma}^{2}-{1\over 4}(q-2)^{2}-l(l+q-2)
\Bigr)\right]\chi(r)=W(r)\chi(r) .
&(3.10)\cr}
$$
At this stage, the Laurent expansions (2.5) and (2.6) lead to
an infinite system of equations for the coefficients $c_{n}$
in the form (cf (2.8))
$$
\Bigr[(n+{\widetilde \gamma})^{2}
-{\widetilde \lambda}^{2}\Bigr]c_{n}
=\sum_{m=-\infty}^{\infty}{\overline u}_{n-m}c_{m}
\eqno (3.11)
$$
where we have defined (cf (3.2))
$$
{\widetilde \lambda}^{2} \equiv l(l+q-2)+{1\over 4}(q-2)^{2}
+w_{0}=\left(l+{1\over 2}(q-2)\right)^{2}+w_{0}
\eqno (3.12)
$$
whereas the notation (2.10) remains unchanged. Thus, one can always
perform the analysis in terms of the infinite matrix (3.4), 
provided that one defines $\widetilde \gamma$ and
${\widetilde \lambda}^{2}$ as in (3.9) and (3.12), respectively. 
The resulting Hill-type equation which leads, in principle, to
the evaluation of the fractional part of the 
polydromy parameter $\gamma$, involves an
even periodic function of $\gamma+{1\over 2}(q-2)$.
\vskip 10cm
\leftline {\bf 4. Asymptotic form of the solutions}
\vskip 0.3cm
\noindent
Since one might be eventually interested in the $S$-matrix, it is
quite important to study the limiting behaviour of stationary
states as $r \rightarrow 0$ and as $r \rightarrow \infty$. In
the former case, one can perform a JWKB analysis of Eq. (1.1),
setting therein
$$
\psi(r)=A(r)e^{{i\over \hbar}S(r)}.
\eqno (4.1)
$$
This leads to the equation (the prime denoting differentiation
with respect to $r$)
$$ 
\eqalignno{
\; & \biggr[\Bigr(2m(E-U(r)) -S'^{2}\Bigr)A
+i{\hbar} \left(2A'S'+AS''
+{(q-1)\over r}AS' \right) \cr
&+{\hbar}^{2}\left(A''+{(q-1)\over r}A'-{l(l+q-2)\over r^{2}}A
\right)\biggr]=0.
&(4.2)\cr}
$$
If, in a first approximation, the term on the second line of
Eq. (4.2) is neglected, one finds the equations
$$
S'^{2}=2m(E-U(r))
\eqno (4.3)
$$
$$
{d\over dr}(A^{2}S')+{(q-1)\over r}A^{2}S' =0
\eqno (4.4)
$$
which imply
$$
S'=\pm \sqrt{2m(E-U(r))}
\eqno (4.5)
$$
$$
A^{2}S'={\rm constant} \; \times r^{-(q-1)}
\eqno (4.6)
$$
and hence, for some constant $\beta$,
$$
A(r)=\beta \; r^{-{(q-1)\over 2}} 
\Bigr(2m(E-U(r))\Bigr)^{-{1\over 4}}.
\eqno (4.7)
$$

To second order in $\hbar$, one has to consider the second line 
of Eq. (4.2). On taking the prefactor $A(r)$ in the form (4.7),
one has then to evaluate the phase $S(r)$ from the equation
$$
S'(r)=\pm \sqrt{2m(E-U(r))+{\hbar}^{2}f_{lq}(A(r))}
\eqno (4.8)
$$
where
$$
\eqalignno{ \; & f_{lq}(A(r)) \equiv {A''\over A}
+{(q-1)\over r}{A' \over A}-{l(l+q-2)\over r^{2}} \cr
&=-\left[{1\over 4}(q^{2}-4q+3)+l(l+q-2)\right]r^{-2} \cr
&+{m\over 2}U'' \Bigr(2m(E-U(r))\Bigr)^{-1}
+{5\over 4}m^{2}U'^{2}\Bigr(2m(E-U(r))\Bigr)^{-2}.
&(4.9)\cr}
$$
Only an approximate calculation of the square root in Eq. (4.8)
is possible, if
$$
\rho \equiv {{\hbar}^{2}f_{lq}(A(r)) \over
2m(E-U(r))} << 1
$$
by expanding $\sqrt{1+\rho}$ in powers of $\rho$, but this does
not improve substantially the understanding of the behaviour
of the wave function as 
$r \rightarrow 0$ for a fixed value of $k$ (see below),
and hence we do not present further calculations along these
lines. One should bear in mind, however, that the JWKB expansion
has an asymptotic nature, and rigorous error bounds 
can be obtained [14].

In particular, in the physically more relevant case of
three spatial dimensions, Eq. (4.7) leads to (see
(1.2) and (1.3))
$$
A(r)={\widetilde \beta} \; r^{-1}
\Bigr(k^{2}-V(r)\Bigr)^{-{1\over 4}}.
\eqno (4.10)
$$
When $r \rightarrow 0$, $V(r)$ is much larger than $k^{2}$
{\it for a fixed value} of $k$, and
hence the JWKB ansatz (4.1) leads to (hereafter $\hbar=1$)
$$
\psi_{I,II} \sim B_{I,II} \; 
r^{-1}(V(r))^{-{1\over 4}} \;
\exp \int_{r}^{r_{0}}\sqrt{V(y)} \; dy
\eqno (4.11)
$$
for some parameters $B_{I,II}$ depending on $\gamma$ and $l$
(cf [6, 8, 18]). Of course, the JWKB solution for {\it all values}
of $k$ which results from (4.5) and (4.10) is, instead,
$$
\psi_{I,II} \sim {\widetilde \beta}_{I,II}r^{-1}
(k^{2}-V(r))^{-{1\over 4}}
\; \exp \left[i \int_{r_{0}}^{r}\sqrt{k^{2}-V(y)} \; dy \right].
\eqno (4.12)
$$

Moreover, as $r \rightarrow \infty$, one has the familiar 
asymptotic behaviour
$$ \eqalignno{
\; & \psi_{I,II} \sim A_{I,II}^{+} \; r^{-1} 
\exp \left \{i \left[kr-l {\pi \over 2}\right] \right \} \cr
&+ A_{I,II}^{-} \; r^{-1}
\exp \left \{-i \left[kr-l {\pi \over 2}\right] \right \}.
&(4.13)\cr}
$$
The $S$-matrix is given by the formula [6, 8]
$$
S={\Bigr(A_{I}^{+}B_{II}-A_{II}^{+}B_{I}\Bigr)
\over \Bigr(A_{I}^{-}B_{II}-A_{II}^{-}B_{I}\Bigr)}
\eqno (4.14)
$$
where the $A$ and $B$ parameters are the ones occurring in the
asymptotic expansions (4.11) and (4.13), and can be obtained by
means of the saddle-point method [6, 8]. 
\vskip 0.3cm
\leftline {\bf 5. Intertwining operators for singular potentials}
\vskip 0.3cm
\noindent
Since exact solutions of singular scattering problems in terms
of special functions are known in a few cases only, it appears
quite important to look for a technique that makes it possible
to generate solutions for complicated problems, relying on what
is known in simpler cases. For this purpose, we here consider
the Darboux method of intertwining operators [19--22]. 

The aim of the Darboux method is to generate families of 
isospectral Hamiltonians. It relies on a theorem which, in modern
language, can be stated as follows [23]. Let $\psi$ be the 
general solution of the Schr\"{o}dinger equation
$$
H \psi(x) \equiv \left[-{d^{2}\over dx^{2}}+V(x) \right]\psi(x)
=E \psi(x) .
\eqno (5.1)
$$
If $\varphi$ is a particular solution of (5.1) corresponding to an
energy eigenvalue $\varepsilon \not = E$, then
$$
{\widetilde \psi}={1\over \varphi}\left(\psi {d\varphi \over dx}
-{d\psi \over dx}\varphi \right)
\eqno (5.2)
$$
is the general solution of the Schr\"{o}dinger equation
$$
{\widetilde H} \; {\widetilde \psi}(x)=E{\widetilde \psi}(x)
\eqno (5.3)
$$
where
$$
{\widetilde H} \equiv -{d^{2}\over dx^{2}}+{\widetilde V}(x)
\eqno (5.4)
$$
$$
{\widetilde V}(x) \equiv V(x)-2{d^{2}\over dx^{2}}\log \varphi(x) .
\eqno (5.5)
$$

In other words, if two Hamiltonian operators, say $H_{A}$ and
$H_{B}$, are given, one looks for a differential operator, say $D$,
such that [24]
$$
H_{B} \; D=D \; H_{A} .
\eqno (5.6)
$$
It is then possible to relate the eigenfunctions of $H_{A}$ and
$H_{B}$ by using the action of $D$ (see below). Here, we focus
on one-dimensional problems, with
$$
H_{A}=H_{0} \equiv -{d^{2}\over dx^{2}}+V_{0}(x)
\eqno (5.7)
$$
$$
H_{B}=H_{1} \equiv -{d^{2}\over dx^{2}}+V_{1}(x)
\eqno (5.8)
$$
$$
D \equiv {d\over dx}+G(x)
\eqno (5.9)
$$
where $V_{0}$ and $V_{1}$ are the ``potential" functions, and $G$
is another function, whose form is determined by imposing the
condition (5.6). This reads, explicitly,
$$
\left(-{d^{2}\over dx^{2}}+V_{1}\right)
\left({d\over dx}+G \right)f=\left({d\over dx}+G \right)
\left(-{d^{2}\over dx^{2}}+V_{0}\right)f
\eqno (5.10)
$$
for all functions $f$ which are at least of class $C^{3}$. On
imposing Eq. (5.10), one finds exact cancellation of the terms
$-{d^{3}f \over dx^{3}}$ and $-G{d^{2}f \over dx^{2}}$, since
they occur on both sides with the same sign. Hence one deals
with the equation
$$
\left[\Bigr(-2G'+V_{1}-V_{0}\Bigr){d\over dx}
+\Bigr(-G''-V_{0}'+(V_{1}-V_{0})G \Bigr)\right]f=0
\eqno (5.11)
$$
which implies
$$
2G'=V_{1}-V_{0}
\eqno (5.12)
$$
$$
-G''-V_{0}'+(V_{1}-V_{0})G=0
\eqno (5.13a)
$$
by virtue of the arbitrariness of $f$. It is now possible to use
Eq. (5.12) to express Eq. (5.13a) in the form
$$
{d\over dx}(-G'+G^{2})={d\over dx}V_{0}
\eqno (5.13b)
$$
which is solved by 
$$
G^{2}-G'=V_{0}+C
\eqno (5.14)
$$
for some constant $C$. Equation (5.14) is known as the Riccati 
equation. Its non-linear nature makes it desirable to develop
an algorithm to relate it, instead, to the solution of a
linear problem. This is indeed achieved by considering the
function $\varphi$ such that
$$
G=-{d\over dx}\log \varphi .
\eqno (5.15)
$$
The Eqs. (5.12) and (5.15) are, of course, in complete agreement
with the result (5.5), with $\widetilde V$ replaced by $V_{1}$,
and $V$ replaced by $V_{0}$.
One then finds, by virtue of (5.14) and (5.15), that $\varphi$
obeys the linear second-order equation
$$
H_{0}\varphi=-C \varphi .
\eqno (5.16)
$$
This is a simple but deep result: one first has to find the 
eigenfunctions of $H_{0}$, say $\varphi$, belonging to the 
eigenvalue $-C$. Once this is achieved, the desired function 
$G$ is obtained from (5.15), and hence the intertwining
operator is
$$
D={d\over dx}-{\varphi ' \over \varphi}.
\eqno (5.17)
$$
In the applications, it is also convenient to use Eq. (5.12)
to express Eq. (5.14) in the form
$$
G^{2}+G'=V_{1}+C.
\eqno (5.18)
$$

If one studies Eq. (1.1) or, in particular, Eq. (2.1), the
standard definition in three dimensions 
$$
\psi(r) \equiv {y(r)\over r}
\eqno (5.19)
$$
leads to a second-order differential operator acting on $y$
which is of the form (5.7) or (5.8). However, the choice of 
a suitable intertwining operator, aimed at relating operators
$H_{A}$ and $H_{B}$ whose potential terms differ in a somehow
substantial way, is a non-trivial task. For example, if one
considers
$$
V_{1}(r) \equiv {A\over r^{4}}+{B\over r^{3}}
\eqno (5.20)
$$
Eq. (5.18) may be then satisfied by (cf (5.9))
$$
G(r)={k\over r^{2}}
\eqno (5.21)
$$
provided that $A=k^{2}$, $B=-2k$ and $C=0$. However, the
resulting potential $V_{0}(r)$ is found to be, from Eq. (5.12),
$$
V_{0}(r)={k^{2}\over r^{4}}+{2k\over r^{3}}
\eqno (5.22)
$$
so that the intertwining operator ends up by relating operators
$H_{A}$ and $H_{B}$ whose potential terms have precisely the
same functional form. A scheme of broader validity, however, is
obtained by looking for $V_{1}(r)$ and $G(r)$ in the
form of (Laurent) series, i.e.
$$
V_{1}(r)=\sum_{n=-\infty}^{\infty}a_{n}r^{n}
\eqno (5.23)
$$
$$
G(r)=\sum_{p=-\infty}^{\infty}b_{p}r^{p}.
\eqno (5.24)
$$
The insertion of (5.23) and (5.24) into Eq. (5.18) leads to
the infinite system
$$
(n+1)b_{n+1}+\sum_{p=-\infty}^{\infty}b_{p}b_{n-p}
=a_{n}+C \delta_{n,0}
\eqno (5.25)
$$
that should be solved, in principle, for $b_{n}$, for
all $n$. One then finds, from Eq. (5.12), a Laurent series
for $V_{0}$ as well, i.e.
$$
V_{0}(r)=\sum_{n=-\infty}^{\infty}f_{n}r^{n}
\eqno (5.26)
$$
where
$$
f_{n}=-(n+1)b_{n+1}+\sum_{p=-\infty}^{\infty}b_{p}b_{n-p}
-C \delta_{n,0}.
\eqno (5.27)
$$

For example, if one takes $V_{1}(r) \equiv g^{2}r^{-4}$,
one has
$$
a_{n}=g^{2}\delta_{n,-4}
\eqno (5.28)
$$
and hence one deals with the infinite system
$$
(n+1)b_{n+1}+\sum_{p=-\infty}^{\infty}b_{p}b_{n-p}
=g^{2}\delta_{n,-4}+C \delta_{n,0} .
\eqno (5.29)
$$
This is a non-linear algebraic system for which it does not seem
possible to obtain a solution such that only a few $b_{p}$ 
coefficients are non-vanishing. For example, if one tries to get
$b_{p}=0$ unless $p=-3,-2,-1$, one finds, on setting
$n=-6,-5,-4,-3,-2,0$ in (5.29) the six equations
$$
(b_{-3})^{2}=0
\eqno (5.30)
$$
$$
2b_{-3}b_{-2}=0
\eqno (5.31)
$$
$$
-3b_{-3}+2b_{-3}b_{-1}+(b_{-2})^{2}=g^{2}
\eqno (5.32)
$$
$$
2(b_{-1}-1)b_{-2}=0
\eqno (5.33)
$$
$$
(b_{-1}-1)b_{-1}=0 
\eqno (5.34)
$$
$$
0=C
\eqno (5.35)
$$
whereas $n=-1$ leads to a trivial identity. Now equations (5.30)
and (5.31) imply that $b_{-3}=b_{-2}=0$, and hence $g^{2}=0$ from
(5.32), which is incompatible with our assumptions. The remaining
equations (5.33)--(5.35) allow for $b_{-1}=1$, further to 
$b_{-1}=0$, but with $C=0$.

However, the implications remain of high interest: to find 
non-trivial solutions with $g^{2} \not =0$ and $C \not = 0$
one needs a large number of $b_{p}$ coefficients (maybe
infinitely many), including those with $p > 0$. This still means
that one has the opportunity to solve the Schr\"{o}dinger 
equation with a complicated singular potential, starting from
what one knows when the potential equals $g^{2}r^{-4}$ [8].
For this purpose, on denoting again by $\varphi$ the eigenfunction
of $H_{0}$ belonging to the eigenvalue $-C$, and by
$\chi \equiv D \varphi$ the eigenfunction of $H_{1}$ belonging to
the same eigenvalue, we notice that the desired $\varphi$ can
be written in the form
$$
\varphi(r)=\int_{0}^{\infty}K(r,r')\chi(r')dr'
\eqno (5.36)
$$
where $K(r,r')$ denotes the Green kernel of the intertwining
operator $D \equiv {d\over dr}+G(r)$. We need such an integral
formula because we have chosen, in our particular example, the
form of the potential term $V_{1}(r)$ in the Hamiltonian $H_{1}$,
for which the scattering states are already known in the
literature [8]. The unknown are instead the scattering states
resulting from the Hamiltonian operator with potential term
equal to $V_{0}(r)$ (see (5.26)).
\vskip 0.3cm
\leftline {\bf 6. Concluding remarks}
\vskip 0.3cm
\noindent
Our paper has studied some aspects of scattering from singular
potentials in quantum mechanics. Its contributions are as follows.
\vskip 0.3cm
\noindent
(i) The technique of Fubini and Stroffolini [6, 8], with emphasis
on the polydromy properties of the wave function, has been applied
to an arbitrary number of spatial dimensions, say $q$, when the
potential admits a Laurent series expansion. The equation obeyed
by the polydromy parameter, $\gamma$, involves a function which
is an even periodic function of $\gamma+{1\over 2}(q-2)$. 
Interestingly, one can rely entirely on the 
analysis performed in [6, 8], provided that one considers the
parameters defined in (3.9) and (3.12) (strictly, the authors
of [6, 8] start from three dimensions, but use a transformation
[9] leading to an equation formally analogous to the radial part
of the stationary Schr\"{o}dinger equation in two dimensions).
\vskip 0.3cm
\noindent
(ii) The Darboux technique of intertwining operators has been
applied to relate the singular potential terms in the
Schr\"{o}dinger equation for stationary states. The
algorithm resulting from Eqs. (5.23)--(5.27) leads, in
particular, to the non-linear algebraic system (5.29) if
the potential $V_{1}$ is taken to be $g^{2}r^{-4}$.

Ultimately, one might want to use these properties to study
quantum field theories which are not perturbatively
renormalizable, according to the original motivations for this
research field [1, 2, 9]. For this purpose, it seems crucial, to
us, to consider the quantum gravity problem, focusing (at least)
on the following questions.
\vskip 0.3cm
\noindent
(i) What is the counterpart, in quantum gravity, of the
Bethe--Salpeter equation containing effective potentials 
of the singular type? As is well known, this equation arises
in the course of studying the quantum theory of relativistic 
bound states, and unfortunately a simple extension of the
Schr\"{o}dinger equation is not available [25]. Even on
neglecting curvature effects due to gravitational fields,
one then faces retardation effects which lead to an extra
relative time variable in the problem [25]. An alternative
description uses a mediating field, whose quantum properties,
however, cannot be ignored [25]. When a quantum theory of
gravity is considered in a space-time approach [26], one may
expect of using the (formal) theory of the effective action,
with a corresponding set of integro-differential equations.
These should be solved, in principle, by using the functional 
calculus. But even if one were able to achieve so much, and
hence derive an effective potential which is a gravitational
counterpart of the potential normally used to reduce the number
of degrees of freedom of relativistic bound-state problems, the
problem would remain of giving a proper interpretation of such
potentials, since they seriously affect the exact theory and
may introduce fictitious singularities (see page 493 of [25]).
\vskip 0.3cm
\noindent
(ii) What kind of results can be then ``imported" on mapping
quantum gravity into a scattering problem from singular 
potentials (e.g. the asymptotic behaviour of the phase shift 
[9, 13], the exact or approximate solutions derived with some
particular choices of singular potentials [1--8, 10, 14],
the existence theorem for the wave operators [12])?
\vskip 0.3cm
\noindent
(iii) How fundamental is the Darboux method of intertwining operators
[19-22] proposed in our section 5? Since also the variable phase
approach to potential scattering relies on an equation
of Riccati type [7], such a question appears to be non-trivial.

The above issues seem to point out that new perspectives are in
sight in the analysis of potential scattering for a wide class
of singular potentials, with possible implications for a
longstanding problem, i.e. the key features of a quantum theory
of the gravitational field (see [16] and references therein). 
Hence we hope that the present work, although devoted to some
technical issues, may lead to a thorough investigation of
quantum gravity from a point of view well grounded in the general
framework of modern high energy physics (cf [27]).
\vskip 0.3cm
\leftline {\bf Acknowledgments}
\vskip 0.3cm
\noindent
The author is much indebted to Giuseppe Marmo for teaching him
all he knows about the Darboux method. The present paper
is dedicated to the memory of Roberto Stroffolini. 
\vskip 0.3cm
\leftline {\bf Appendix}
\vskip 0.3cm
\noindent
To be self-contained, let us describe what is meant by Fuchsian
singularities of second-order differential equations. A theorem
due to the German mathematician Immanuel Lazarus Fuchs states that
a necessary and sufficient condition for the linear equation
$$
\left[{d^{2}\over dx^{2}}+p_{1}(x){d\over dx}+p_{2}(x)
\right]y(x)=0
\eqno (A1)
$$
to admit a fundamental system of integrals, say $y_{1}(x)$ and
$y_{2}(x)$, which, in the neighbourhood of the singular point
$x_{0}$, can be expressed as [with $\varphi_{2}, \varphi_{2},
\psi$ analytic functions in the neighbourhood of $x_{0}$, for
some constants $r_{1},r_{2},{\tilde r}_{1},A: (r_{1}-r_{2})
\not \in Z$]
$$
y_{1}(x)=(x-x_{0})^{r_{1}}\varphi_{1}(x)
\eqno (A2a)
$$
$$
y_{2}(x)=(x-x_{0})^{r_{2}}\varphi_{2}(x)
\eqno (A2b)
$$
or
$$
y_{1}(x)=(x-x_{0})^{{\tilde r}_{1}} \varphi_{1}(x)
\eqno (A3a)
$$
$$
y_{2}(x)=y_{1}(x)\Bigr[A\log(x-x_{0})+\psi(x)\Bigr]
\eqno (A3b)
$$
is that $p_{1}$ and $p_{2}$ should have poles of order
$\leq 1$ and $\leq 2$, respectively, at the singular point
$x_{0}$. One then says that $x_{0}$ is a {\it Fuchsian singularity}
for equation (A1).

To study the point at infinity, one defines
$$
\xi \equiv {1\over x}
\eqno (A4)
$$
which leads to the equation (cf (A1))
$$ 
\left \{ {d^{2}\over d \xi^{2}}
+ \left[{2\over \xi}-{1\over \xi^{2}}p_{1}
\left({1\over \xi}\right)\right]{d\over d\xi}
+{1\over \xi^{4}}p_{2}\left({1\over \xi}\right)
\right \}y(\xi)=0.
\eqno (A5)
$$
In the analysis of equation (A5) as $\xi \rightarrow 0$, which
corresponds to the point at infinity of (A1), one can thus use
again the Fuchs theorem, which implies that 
$p_{1}\left({1\over \xi}\right)$ and $p_{2}\left({1\over \xi}
\right)$ should have zeros of degree $\geq 1$ and $\geq 2$,
respectively, at $\xi=0$. If this condition is fulfilled,
equation (A1) is said to be {\it Fuchsian at infinity}.

When all singular points are Fuchsian, the corresponding
differential equation is said to be {\it totally Fuchsian}.
The {\it non-Fuchsian singularities} are, by contrast, singular
points of (A1) for which the above conditions on poles and zeros
of the functions $p_{1}$ and $p_{2}$ are not fulfilled.

In our paper, the word {\it polydromy} refers to the well known
property of some functions of complex variable of being 
multi-valued functions of the independent variable. For example,
if $z$ is complex, its logarithm is given by the formula
$$
\log(z)=\log |z| + i \; {\rm arg}(z).
\eqno (A6)
$$
\vskip 0.3cm
\leftline {\bf References}
\vskip 0.3cm
\item {[1]}
Bastai A, Bertocchi L, Fubini S, Furlan G and Tonin M 1963
{\it Nuovo Cim.} {\bf 30} 1512
\item {[2]}
Bastai A, Bertocchi L, Furlan G and Tonin M 1963 {\it Nuovo Cim.}
{\bf 30} 1532
\item {[3]}
Bertocchi L, Fubini S and Furlan G 1965 {\it Nuovo Cim.}
{\bf 35} 599
\item {[4]}
Bertocchi L, Fubini S and Furlan G 1965 {\it Nuovo Cim.}
{\bf 35} 633
\item {[5]}
De Alfaro V and Regge T 1965 {\it Potential Scattering}
(Amsterdam: North-Holland)
\item {[6]}
Fubini S and Stroffolini R 1965 {\it Nuovo Cim.} 
{\bf 37} 1812
\item {[7]}
Calogero F 1967 {\it Variable Phase Approach to Potential
Scattering} (New York: Academic Press)
\item {[8]}
Stroffolini R 1971 {\it Nuovo Cim.} {\bf A 2} 793
\item {[9]}
Frank W M, Land D J and Spector R M 1971 {\it Rev. Mod. Phys.}
{\bf 43} 36
\item {[10]}
Aly H H, M\"{u}ller--Kirsten H J W and Vahedi--Faridi N 1975
{\it J. Math. Phys.} {\bf 16} 961
\item {[11]}
Gribov V A 1976 {\it Theor. Math. Phys.} {\bf 26} 48
\item {[12]}
Enss V 1979 {\it Ann. Phys. (N.Y.)} {\bf 119} 117
\item {[13]}
Dolinszky T 1980 {\it Nucl. Phys.} {\bf A 338} 495
\item {[14]}
Brander O 1981 {\it J. Math. Phys.} {\bf 22} 1229
\item {[15]}
Amrein W O and Pearson D B 1997 {\it J. Phys.} 
{\bf A 30} 5361
\item {[16]}
Esposito G, Kamenshchik A Yu and Pollifrone G 1997 
{\it Euclidean Quantum Gravity on Manifolds with Boundary}
({\it Fundamental Theories of Physics 85})
(Dordrecht: Kluwer)
\item {[17]}
Reed M and Simon B 1975 {\it Methods of Modern Mathematical
Physics. II. Fourier Analysis and Self-Adjointness}
(New York: Academic)
\item {[18]}
Newton R G 1967 {\it Scattering Theory of Waves and Particles}
(New York: McGraw Hill)
\item {[19]}
Darboux G 1882 {\it C. R. Acad. Sci. (Paris)} {\bf 94} 1456
\item {[20]}
Ince E L 1956 {\it Ordinary Differential Equations}
(New York: Dover)
\item {[21]}
Deift P A 1978 {\it Duke Math. J.} {\bf 45} 267
\item {[22]}
Carinena J F, Marmo G, Perelomov A M and Ranada M F 1998
Related operators and exact solutions of Schr\"{o}dinger
equations {\it Preprint}
\item {[23]}
Luban M and Pursey D L 1986 {\it Phys. Rev.} 
{\bf D 33} 431
\item {[24]}
Esposito G and Marmo G 1998 {\it From Classical to Quantum
Mechanics} (in preparation)
\item {[25]}
Itzykson C and Zuber J B 1985 {\it Quantum Field Theory}
(New York: McGraw--Hill)
\item {[26]}
DeWitt B S 1984 in {\it Relativity, Groups and Topology II}
eds B S DeWitt and R Stora (Amsterdam: North Holland)
\item {[27]}
Ashtekar A 1987 {\it Asymptotic Quantization}
(Naples: Bibliopolis)

\bye